\documentclass[twocolumn,groupedaddress,longbibliography]{revtex4-1}
\usepackage[utf8]{inputenc}
\usepackage[T1]{fontenc}
\usepackage{graphicx, placeins}
\usepackage[unicode=true,pdfusetitle,bookmarks=true,bookmarksnumbered=true,bookmarksopen=false, breaklinks=false,pdfborder={0 0 0},pdfborderstyle={},backref=false,colorlinks=false, pdftitle={Single-exposure absorption imaging of ultracold atoms using deep learning}]{hyperref}
\begin{document}

\title{Single-exposure absorption imaging of ultracold atoms using deep learning}

\author{Gal Ness, Anastasiya Vainbaum, Constantine Shkedrov, Yanay Florshaim, Yoav Sagi}
\email[Electronic address: ]{yoavsagi@technion.ac.il}

\affiliation{Physics Department and Solid State Institute, Technion -- Israel Institute of Technology, Haifa 32000, Israel}

\date{\today}
\begin{abstract}
Absorption imaging is the most common probing technique in experiments with ultracold atoms. The standard procedure involves the division of two frames acquired at successive exposures, one with the atomic absorption signal and one without. A well-known problem is the presence of residual structured noise in the final image, due to small differences between the imaging light in the two exposures. Here we solve this problem by performing absorption imaging with only a single exposure, where instead of a second exposure the reference frame is generated by an unsupervised image-completion autoencoder neural network. The network is trained on images without absorption signal such that it can infer the noise overlaying the atomic signal based only on the information in the region encircling the signal. We demonstrate our approach on data captured with a quantum degenerate Fermi gas. The average residual noise in the resulting images is below that of the standard double-shot technique. Our method simplifies the experimental sequence, reduces the hardware requirements, and can improve the accuracy of extracted physical observables. The trained network and its generating scripts are available as an open-source repository (\href{http://absDL.github.io/}{absDL.github.io}).
\end{abstract}

\maketitle

\section{Introduction}

Ultracold atomic gases are unique systems that allow studying few- and many-body physics in a highly precise and tunable manner. The atomic ensembles are exquisitely isolated from the surroundings as they are held in an ultra-high vacuum environment; therefore, probing them is almost always restricted to the analysis of their optical response. The most widely used probing technique is absorption imaging, where a collimated resonant laser beam is passed through the cloud, and the shadow cast by the atoms is recorded by a digital camera \cite{MakingProbing}. The spatial atomic distribution is then extracted from the position-dependent absorption coefficient. The coherence length of the probe beam is typically much longer than the distances between optical interfaces in the experiment, hence, unwanted reflections interfere and generate a characteristic patterns of stripes and Newton's rings in the recorded image. These patterns pose a problem in distinguishing between the signal and the non-uniform background.

\begin{figure*}
\centering
\includegraphics[width=\linewidth]{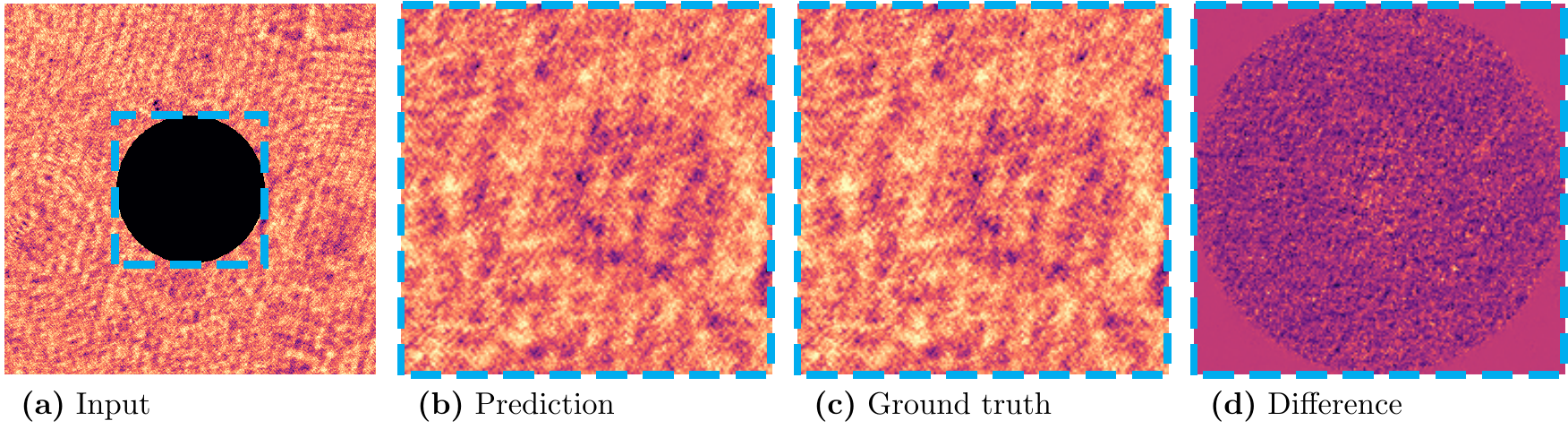}
\caption{Completion of a background frame by the neural net -- an example of network evaluation of a typical image without atoms from the validation set.
	\textbf{(a)} The input log image with its central part masked. The network task is to complete the image in the central cyan square.
	\textbf{(b)} The network prediction for the central square.
	\textbf{(c)} The original central part of the image (``ground truth'').
	\textbf{(d)} The difference between the network prediction and the ground truth, multiplied by $5$ to enhance the contrast. The residual OD root mean squared error of this example is $0.061$ for both the single-exposure and double-exposure techniques.
}
\label{fig:no_atoms}
\end{figure*}

The standard solution is to employ a double-exposure scheme: the first exposure is performed while the atoms are present, while the second \emph{reference} exposure is performed shortly after and without the atoms. The exposure without atoms can be done either by waiting for the atoms to move out of the frame or by optically pumping them into a dark state. The line-of-sight integrated optical density (OD) image is formed by subtracting the logarithms of the pixel counts in the two frames, with and without the atoms. However, due to acoustic noises and other dynamical processes, the noise patterns in the two images are typically not identical. This results in a residual structured noise pattern in the final image (Fig.\,\ref{fig:no_atoms}a). The lower signal to noise ratio afflicted by the fringes is particularly problematic in low-OD images. Linear approaches for background completion were recently suggested \cite{Niu2018,Song2020}, but, as we show, they are sensitive to small changes in the noise pattern that evolve over time.

In this work, we tackle the noisy background problem using machine learning, a term describing a set of algorithms that effectively perform a specific tasks relying on patterns and inference. Among these, deep learning refers to a class of models which involves information propagation via multiple structures, enabling the translation of a given input to a certain prediction. The use of deep learning has become widespread in recent years for problems where an analytic mapping does not exist or when numeric solutions are intractable \cite{LeCun2015,Biamonte2017,Mehta2019,Carleo2019,Raghu2020}. Image completion is an excellent example of such an application, particularly in a scenario where there are typical recurrent but varying patterns in the image. Machine learning techniques were also used for the optimization of ultracold atoms cooling sequences \cite{Wigley2016,Tranter2018,Nakamura2019,Barker2019} and to execute related numerical calculations \cite{Pilati2019}. They were also suggested \cite{Picard2019} and demonstrated \cite{Ding2019} to be useful for fluorescence detection of pinned atoms and ions. 

Here we report on a new approach for absorption imaging that uses a deep neural network (DNN) to generate an ideal reference frame from a single image that includes the atomic absorption signal. The reference image is constructed by masking out the part of the image containing the atomic shadow and using the network for image completion of the background. We demonstrate the new method with data acquired with ultracold $^{40}\mathrm{K}$ gas and show that the images captured by the single exposure technique feature lower noise levels and therefore allow for a more accurate extraction of physical observables. In addition to the improvement in the data quality, our single-shot approach simplifies the experimental sequence and eases the hardware requirements from the camera. The DNN model successfully adapts to both short and long time variations, and therefore it constitutes a robust solution.

\section{Methods}

\paragraph*{Experimental apparatus.}

The experiments are conducted with a quantum degenerate Fermi gas of $^{40}\mathrm{K}$ atoms with an equal mixture of the two lowest energy states in the $F=9/2$ manifold at a magnetic field of $185$G. Our experimental system and cooling procedure are the same as described in Refs.\,\cite{Shkedrov2018,Ness2018}. The frames without atoms were captured deliberately along seven months to test the DNN in realistic conditions. We acquired data with atomic clouds at different conditions by modifying the evaporation cooling sequence. For training and validation of the DNN, we also acquired images without atoms. To this end, we set the initial position of the optical transfer trap to about $2\mathrm{cm}$ away from their location at the magnetic trap, hence no atoms are shuttled to the position where the images are recorded. In all cases, the first exposure was taken between $12\mathrm{ms}-18\mathrm{ms}$ after the optical dipole trap was turned off abruptly.

The images are taken with a laser tuned to the cycling transition $F=9/2,m_F=-9/2\rightarrow F'=11/2,m_F=-11/2$ in the $D_2$ manifold, at a wavelength of $\sim766.7$nm. The laser linewidth is about $100\mathrm{kHz}$, much narrower than the $D_2$ natural linewidth of $\sim2\pi\times 6$MHz. The illumination is pulsed for  $80\mathrm{\mu s}$ and recorded by a $14$ bit CCD camera \cite{pixelflyModel}. The reference frame (for the conventional absorption imaging) is recorded with a second pulse given after $50\mathrm{ms}$, when the atoms already moved out of the camera field of view. We also capture ``dark frames'' without illumination at all that serve as the zero references. The dark images don't have to be taken often since they only account for any remaining light which is not due to the probe beam and for electronic noise in the camera. Prior to analyzing the two images in the conventional absorption imaging technique, we correct for small differences which may exist between the intensity of the illumination in both exposures. These differences are typically of few percents. The second exposure is taken only in order to compare our technique with the conventional method and is neither required for the application of the DNN nor for its training.

Two physical observables that are commonly used in ultracold atomic experiments are the temperature and number of atoms. In the presented results, the number of atoms in the cloud and its temperature are controlled by changing the final trap depth in the optical evaporation. We extract the observables from the momentum distribution, which is measured after $15\mathrm{ms}$ of a ballistic expansion. To extract the observables, we fit the OD images with \cite{MakingProbing}
\begin{equation}\label{eq:FD_dist_2d}
    \mathrm{OD}\left(x,y\right)=\mathrm{OD_{peak}}\frac{Li_2\left(-z e^{-\frac{\left(x-x_0\right)^2}{2\sigma_x^2}-\frac{\left(y-y_0\right)^2}{2\sigma_y^2}}\right)}{Li_2\left(-z\right)}+B\;, 
\end{equation}
where $\mathrm{Li}_{n}\left(z\right)$ denotes the Jonqui\`{e}re's polylogarithm function, $z=e^{\mu/k_BT}$ is the fugacity, and $B$ accounts for any remaining constant background in the OD image. From the fugacity, we extract the relative temperature $T/T_F=\left[-6Li_3\left(-z\right)\right]^{-1/3}$, with $T_F=\left(6N\right)^{1/3}\hbar\bar{\omega}$ being the Fermi temperature, and $\bar{\omega}$ is the geometrically-averaged trapping frequency, which we measure and rescale according to the trapping laser power. The number of atoms, $N$, is obtained by integrating over the fitted momentum distribution.

\paragraph*{DNN architecture and training.}
DNNs establish a pipeline where the input (the information in the masked OD image, in our case) undergoes multiple convolutional transformations and dimensional variations. These transformations distill the features of the underlying spatial pattern, and their result is the prediction of the DNN. The network is trained to optimally recover the structure of the illumination in the region where the atomic signal appears. The training phase is performed using images captured \emph{without atoms}, and constitute therefore the ``ground truth'' for the unsupervised reconstruction. At each optimization step, the prediction of the network is compared to the ground truth values in the masked area, and the weights of the model are varied to minimize the loss, i.e., the mean squared error ($L_2$ norm) between the ground truth and the prediction.
At the end of the training, we obtain an optimized model ready for prediction (inference) on new images \emph{with atoms}. The network produces an ideal reference regardless of whether atoms appeared in the original image or not, because the relevant region is masked out. Since the involved convolutions are relatively simple, the evaluation of the model for inference on new inputs is rapid, and therefore the integration of a trained network into the infrastructure of another calculation is extremely facile.

From the raw images we subtract the dark frames, and then take the logarithm of their pixel values. The convolutional network is an autoencoder of a U-net architecture \cite{Ronneberger2015}. The input to the network is the OD image cropped to $476\times 476$ pixels around the position of the atoms, from which we mask out the central circle with a diameter of $190$ pixels \cite{calibration} that may include an absorption signal if atoms are present. This mask diameter is larger by at least a factor of two relative to the size of the typical atomic cloud, to ensure that there is no absorption signal in the region used by the DNN to predict the background.
For training, we use a generator to riffle through the stored TIFF images, apply the mask on the input, and feed the DNN input with $8$ frames batches \cite{chollet2015keras}. To evaluate the DNN on an atomic frame, we store the model inference as binary file and subtract the input frame to obtain the atomic OD.
By minimizing the loss over the square circumscribing the masked region (dashed cyan square in Fig.\,\ref{fig:no_atoms}a), we ensure continuity at the corners, where the background is unmasked. Effectively $1-\frac{\pi}{4}$ of the loss is dedicated to image duplication rather than completion, in order to eliminate any offset between the input and output frames, which might be translated into an error in the number of atoms.

The feed-forward network consists of about $20\cdot10^6$ parameters arranged in $27$ layers. These parameters were optimized by running over $\sim30\cdot10^3$ frames captured without atoms, with additional $\sim7\cdot10^3$ images for loss validation, comparing the network output to the original central part of each image, and minimizing the mean squared error loss function. 
We used ADAM optimizer \cite{Kingma2014} and Glorot initialization \cite{glorot2010understanding} for the parameters optimization, applying $99\%$ batch normalization \cite{Ioffe2015}.
For this application, labeling of the input frames is unnecessary as the network output is compared directly against its input before masking. The only prior knowledge is the absence of atoms in the peripheral region and, only for the training set, also in the central area.
Notably, generative adversarial networks \cite{Goodfellow2014}, which were found very successful in natural-scene image competition tasks, might be destructive for this study case, as there is a given unique ground truth.

\section{Results}

\begin{figure}
\centering
\includegraphics[width=\linewidth]{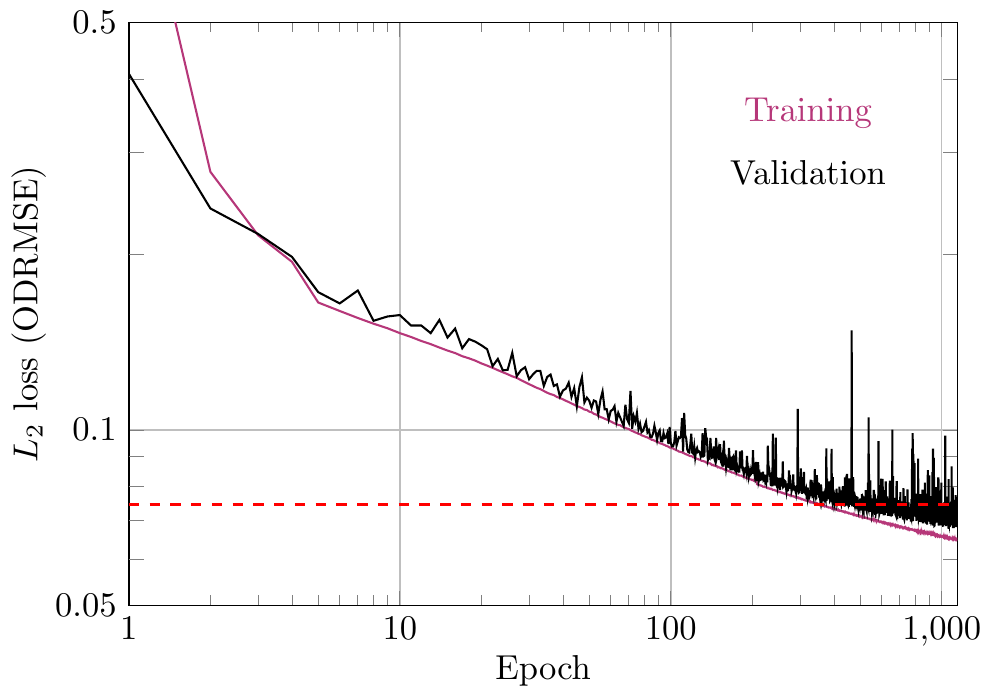}
\caption{Minimization of the residual error along the DNN training. Optical-density root mean squared error between the model prediction and the ground truth as a function of the number of training iterations (epochs). Lower values mean better performance. The purple curve represents the residual loss of the training set, which is minimized in the optimization process. The black curve is the residual error on the validation set, which was not used for training. The dashed red line designates the mean residual noise in the standard double-shot scheme, multiplied by $\pi/4$ to correctly compare with the residual noise of the DNN prediction in the central circle (see Fig.\,\ref{fig:no_atoms}d).
}
\label{fig:loss}
\end{figure}

\paragraph*{DNN performance on the validation set.}

First, we examine the residual noise in inferences on the validation set, which was not used for training and does not include atomic signal. The convergence of the model is depicted in Fig.\,\ref{fig:loss}, where we present the decay of the residual loss during the training process for both the training (purple) and validation (black) datasets. The decay in both datasets on a log-log scale is sub-power-law. It exceeds the reference level, set by the average double-shot residual noise (dashed red line), after approximately $100$ training epochs, which mainly points to a reliable extraction of the bias, but noise features still exist.
In principle, the training should continue as long as the validation loss decreases. In practice, the loss decay slows dramatically after few hundreds of epochs, and we therefore cease the training after $1133$ epochs. An example for image completion without atoms is displayed in Fig.\,\ref{fig:no_atoms}, with the DNN input (\ref{fig:no_atoms}a) and the corresponding prediction of the network (\ref{fig:no_atoms}b), which closely resembles the original data (\ref{fig:no_atoms}c). Notably, there are no significant spatial correlations in the difference between the desired and the predicted frame (\ref{fig:no_atoms}d).

The lowest residual error is $0.0681$ optical-depth root mean squared error (ODRMSE), for the whole validation dataset captured intermittently along seven months. As most of the residual error resulted from the inner circle of the square output image (see Fig.\,\ref{fig:no_atoms}d), a fair comparison for the loss is against $\pi/4$ of the averaged-double-shot error, indicated by a dashed red line in Fig.\,\ref{fig:loss}. This reference value is $0.0745$ (ODRMSE), $9.4\%$ higher than the minimal validation loss obtained during the first $1139$ epochs.

\begin{figure}
\centering
\includegraphics[width=\linewidth]{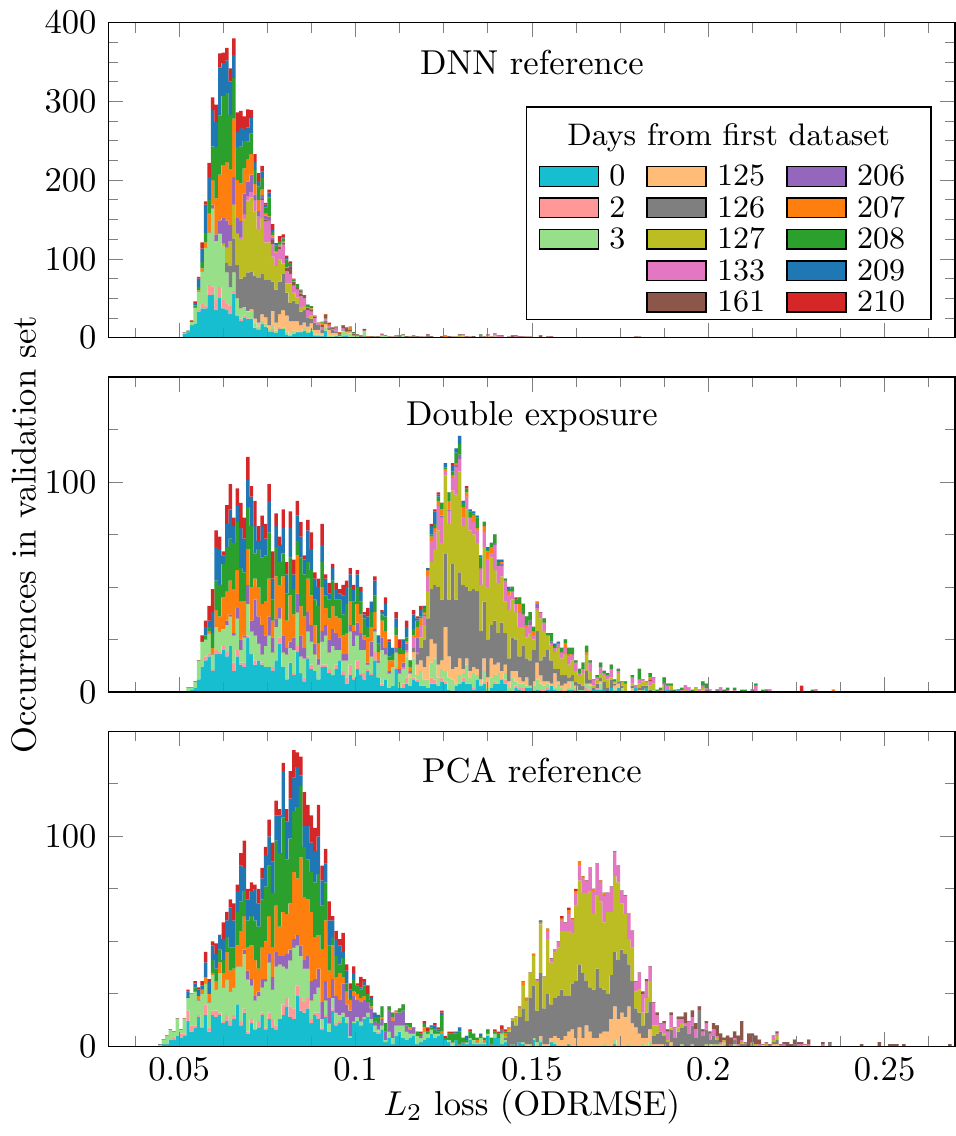}
\caption{Residual error distribution of the difference images in the validation set. The upper histogram indicates the optical-density root mean squared error of the DNN single-exposure technique following $1133$ training epochs. The middle histogram represents the residual error in the standard double-exposure technique, after correction for probe intensity fluctuations. The lower histogram depicts the residual error of PCA-based reference generation images \cite{Niu2018}. The PCA vectors set was extracted from the $300$ significant components out of $600$ random images taken from the DNN training set. Different colors distinguish the validation set constituent frames by date, counting from the first partial set.
}
\label{fig:loss_hist}
\end{figure}

\begin{figure*}
\centering
\includegraphics[width=\linewidth]{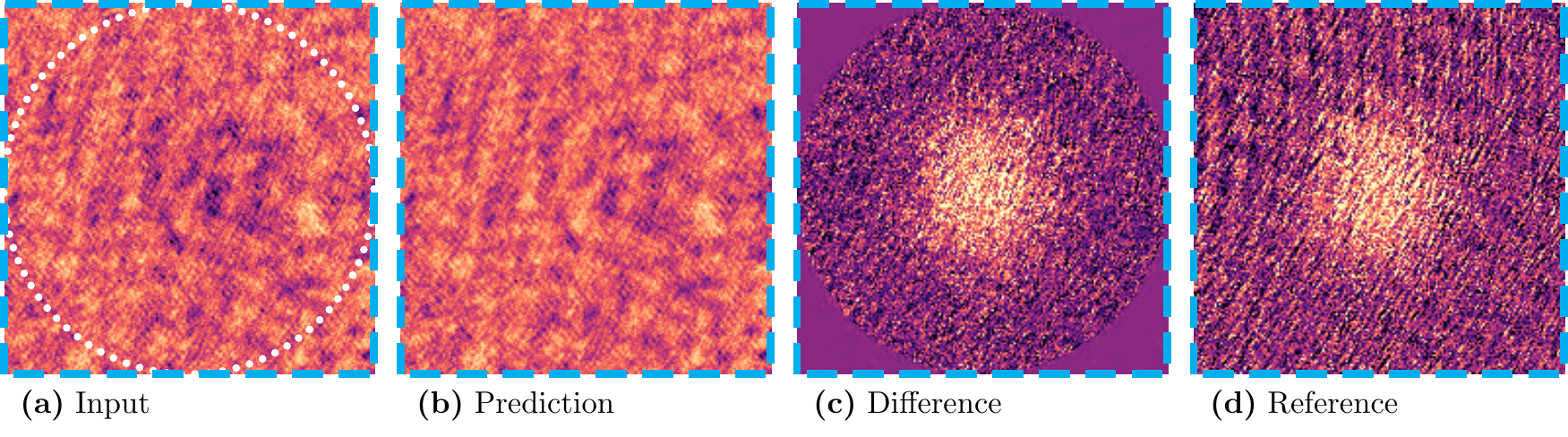}
\caption{Reconstruction of a single-shot image with atoms, exemplified with a cloud of $\sim30\cdot10^3$ atoms.
	\textbf{(a)} The central square of a single-shot log image, including the masked area (dotted white circle).
	\textbf{(b)} The network prediction.
	\textbf{(c)} The difference between prediction and input, multiplied by $5$, resulting in a fringes-free single-shot absorption image.
	\textbf{(d)} The result of the conventional two-exposures technique in the same experiment, where the second exposure is taken $50\mathrm{ms}$ after the first one (also multiplied by $5$).
}
\label{fig:with_atoms}
\end{figure*}

In Fig.\,\ref{fig:loss_hist} we compare the histograms of the residual loss on the validation set using the DNN-based single-shot technique (upper panel), the conventional double-exposure technique (middle panel), and background completion using principal component analysis (PCA) technique (lower panel) \cite{Niu2018}. The histogram for the DNN technique exhibits a single narrow peak, while for the two other approaches it is markedly wider and multi-structured. To illustrate the source of this behavior, we color the histograms based on the elapsed time when taking the corresponding dataset, relative to the first set. We find that the double-peak structure of the conventional double-shot technique is correlated to time variations, probably due to slow drifts in the probe light intensity. An exacerbation of this variation is observable in the PCA results. We substantiate that it directly emerges from the variations in the set from which the PCA basis is taken by repeating the PCA analysis but with the basis taken over $600$ frames all from the first day of image acquisition. In this case, we find that the PCA approach yields excellent results for same-day frames, $0.08(2)$ ODRMSE. Nonetheless, its performance dramatically deteriorates with long-term drifts -- we find distinct date-dependent peaks in the histogram (not shown in the figure), and for the $206-210$ days datasets the error distribution lies at $0.42(1)$ ODRMSE. We can conclude that in order for the PCA approach to maintain adequate performance, recurrent dataset accumulation and analysis is needed, almost on a daily basis. In contrast, the DNN technique is robust and insensitive to these variations. It derives its robustness from the variance in the substantially broader dataset, which is tractable due to the sequential training of the network.

The results on the validation set show that the DNN single-exposure approach achieves lower residual noise levels and deals better with variations in the imaging conditions when compared to the conventional double-shot scheme or linear algorithms. The residual noise of the DNN technique can, in principle, be further reduced by additional training. To assess the usefulness of the time invested in such prolonged training, one should take into account whether it has a measurable effect on physical observables, as we describe in the next section.

\begin{figure*}
\centering
\includegraphics[width=\linewidth]{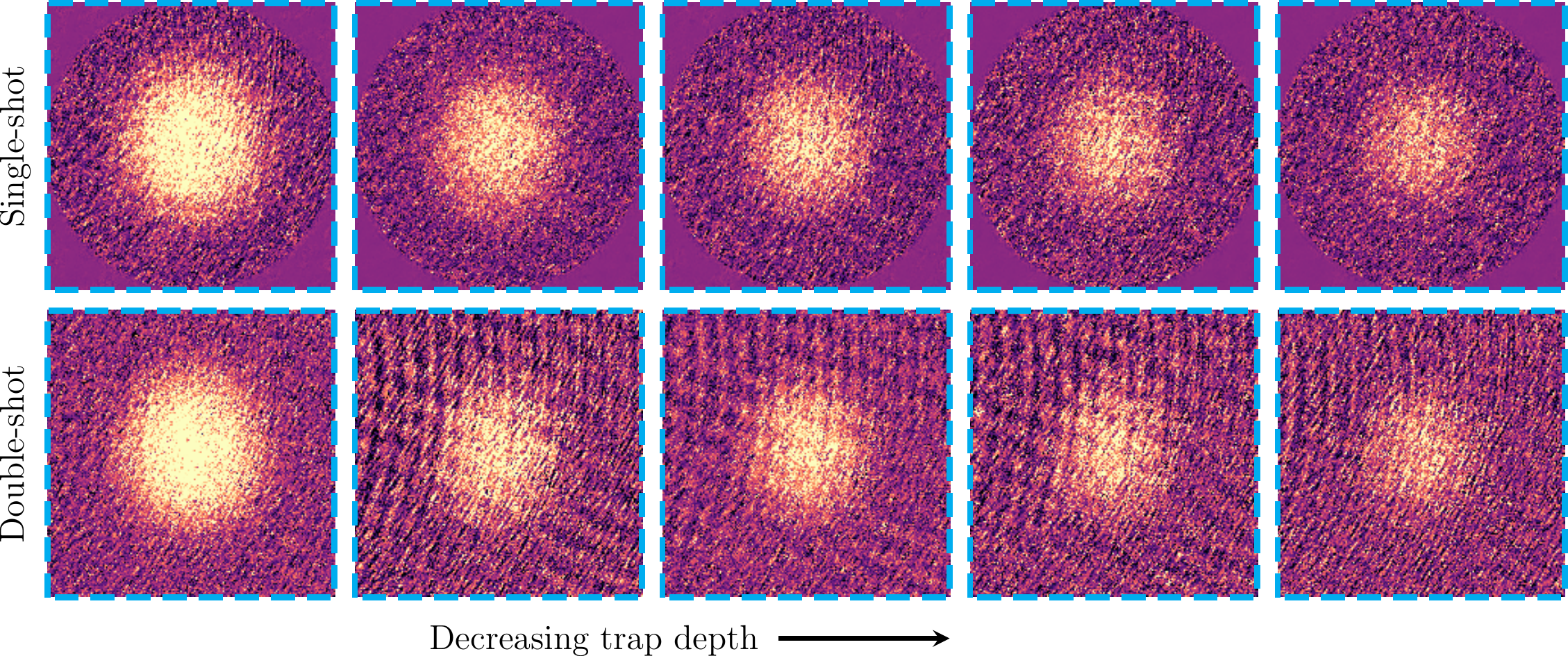}
\caption{Additional examples of inferences of the neural network on images with atoms (upper panel) for different conditions of the atomic cloud. Lower panel presents the correlative results using the standard double-shot technique. The numbers of atoms in these examples are, from left to right, $99(11)\cdot10^3$, $55(7)\cdot10^3$, $45(9)\cdot10^3$, $37(9)\cdot10^3$, and $28(7)\cdot10^3$; and they were released respectively from $190$, $117$, $89$, $76$, and $57\mathrm{nK}$-deep traps. All examples displayed in the same color scale as in Fig.\,\ref{fig:with_atoms}d.
	}
	\label{fig:with_atoms_more_exps}
\end{figure*}

\begin{figure*}
\centering
\includegraphics[width=\linewidth]{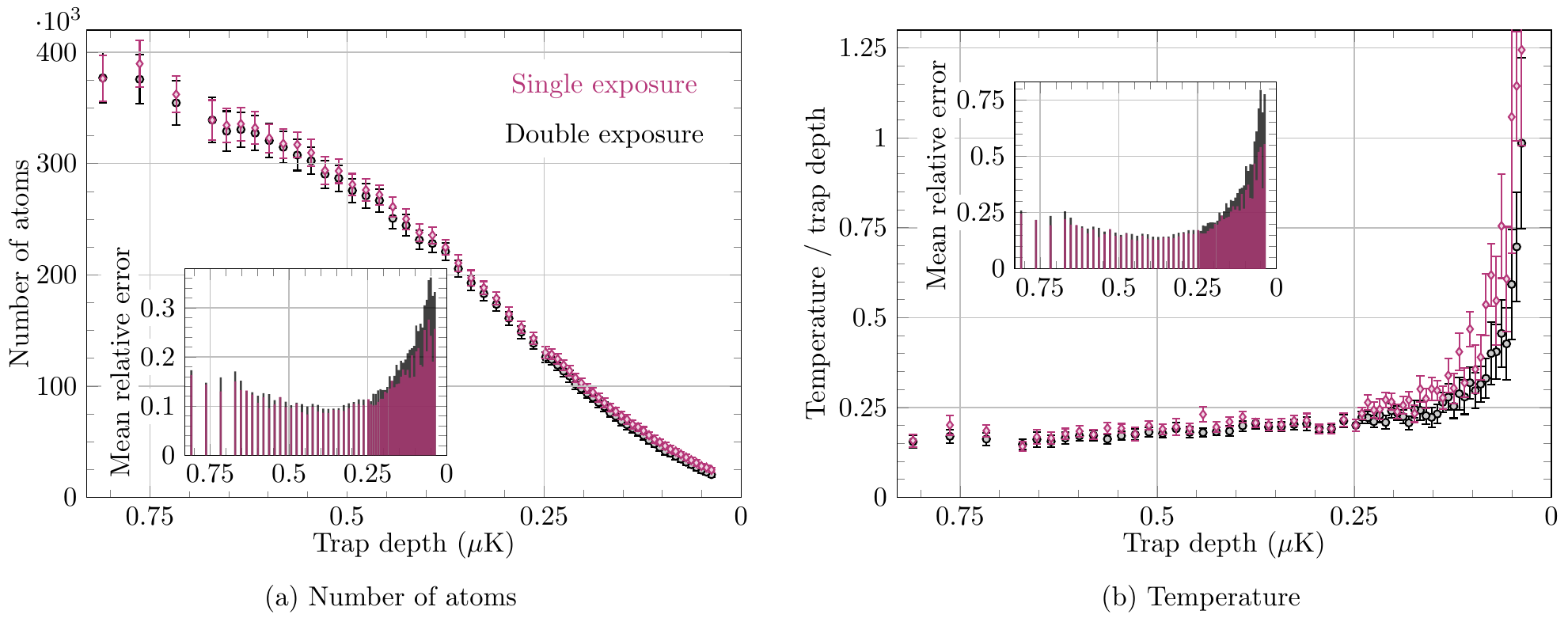}
\caption{Characterization of resulted images -- number of atoms and temperature extracted by fitting a Fermi-Dirac distribution to the data. The conditions of the atomic clouds are controlled by the final trap depth in the optical evaporation. Black circles mark the results of the conventional double-exposure technique, while purple diamonds mark the results with the single-shot DNN approach. Errorbars combine extraction uncertainty with shot-to-shot variation over $10$ experimental realizations. The insets show the average fit extraction error. The single-exposure technique achieves a better accuracy in both observables.}
\label{fig:fitting_results}
\end{figure*}

\paragraph*{Single-shot imaging evaluation.}

In this section we present single-shot absorption images of a quantum degenerate fermionic potassium gas at different conditions. 
A typical analysis of a low-OD image following a ballistic expansion from a $\sim80\mathrm{nK}$-deep trap is shown in Fig.\,\ref{fig:with_atoms}. In panel (\ref{fig:with_atoms}a), we present the inner square part of the input log image. In this example, there are approximately $30\cdot10^3$ atoms, hence the atomic signal is hardly discernible from the background to the naked eye. When it is subtracted from the network prediction in (\ref{fig:with_atoms}b), a clean OD image is obtained (\ref{fig:with_atoms}c). As a comparison, panel (\ref{fig:with_atoms}d) shows the conventional absorption image obtained from two exposures in the same experiment. Evidently, the single-shot approach eliminates the remaining fringe pattern and yields an overall better OD image. More examples for different trap depths are presented in the upper panel of Fig.\,\ref{fig:with_atoms_more_exps}, and show the same behaviour regardless of the atomic conditions.

\paragraph*{Effect on physical observables.}

In Fig.\,\ref{fig:fitting_results} we plot the number of atoms and temperature for different trap depths as extracted by the single-shot (purple diamonds) and the two-exposures (black circles) techniques. Importantly, the new technique does not introduce any systematic error in extraction of these important observables. The errorbars represent the $1\sigma$ shot-to-shot variation in the experimental conditions combined with the fitting extraction error. Since both of these terms are of a similar magnitude, it is hard to observe the improvement in the single-exposure technique. To emphasize this improvement, we present in the insets only the fitting extraction relative error averaged over the $10$ experimental realizations in each trap depth. We find that the extraction uncertainty of both observables is smaller by $\sim17\%$ using the single-exposure technique.

\section{Summary and outlook}

We have demonstrated a single-shot absorption imaging based on a deep convolutional network background completion. We have shown that this approach can accurately reconstruct atomic density profiles and yield smaller errors on the extracted physical quantities, compared to the standard double-exposure technique. The single-shot imaging lifts the need for fast cameras and facilitates multi-framed acquisitions. The corresponding simplification directly enables simpler and cleaner designs for new cold atomic systems. We have also demonstrated the ability of the DNN to adapt to variations in the working condition that develop through time.

Our network can be improved in several aspects. First, the masked area can be enlarged to achieve even better robustness. Also, by training the network over random patches in the uncropped OD image, the position-dependency of the result can be further reduced.
Another interesting direction is the implementation of an \emph{online learning} scheme, where images are routinely added to the dataset and the model is continuously updated between inferences.

\section*{Data availability}

The trained network and its generating scripts are publicly available as an open-source Python software package \cite{absDL} to facilitate their deployment by other experimental groups. Using the provided repository, single-shot imaging can be realized on any imaging apparatus, following local parameters training.
 k

\section*{Acknowledgments}

This research was supported by the Israel Science Foundation (ISF) grant  No.\,1779/19, and by the United States-Israel Binational Science Foundation (BSF), Jerusalem, Israel, grant No.\,2018264. The GeForce TITAN V used for the local network training was donated by the Nvidia Corporation. Remote training power was granted by the Google Cloud Platform research credits program. G.N. would like to thank Amit Oved for inspiring discussions.

\bibliography{bibfile}

\end{document}